\documentclass[11pt,preprint]{aastex}
\usepackage{latexsym}
\usepackage{longtable}

\begin{document}

\begin{center}
IMPROVED LABORATORY TRANSITION PROBABILITIES FOR Er \textsc{ii} AND 
APPLICATION TO THE ERBIUM ABUNDANCES OF THE SUN AND FIVE $r-$PROCESS RICH, 
METAL-POOR STARS 
\end{center}

\begin{center}
(short title: Er \textsc{ii} Transition Probabilities and Abundances)
\end{center}

\begin{center}
J. E. Lawler$^{1}$, C. Sneden$^{2}$, J. J. Cowan$^{3}$, J.-F. Wyart$^{4}$, 
I. I. Ivans$^{5}$, J. S. Sobeck$^{2}$, M. H. Stockett$^{1}$, and E. A. Den 
Hartog$^{1}$
\end{center}

$^{1}$Department of Physics, University of Wisconsin, Madison, WI 53706; 
\underline {jelawler@wisc.edu}, \underline {stockett@wisc.edu}, \underline 
{eadenhar@wisc.edu}

$^{2}$Department of Astronomy and McDonald Observatory, University of Texas, 
Austin, TX 78712; \underline {chris@verdi.as.utexas.edu}, \underline 
{jsobeck@astro.as.utexas.edu}

$^{3}$Department of Physics and Astronomy, University of Oklahoma, Norman, 
OK 73019; \underline {cowan@nhn.ou.edu}

$^{4}$Laboratoire Aim\'{e} Cotton, Centre National de la Recherche 
Scientifique (UPR3321), 91405-Orsay, France; \underline 
{Jean-Francois.Wyart@lac.u-psud.fr}

$^{5}$The Observatories of the Carnegie Institution of Washington, 813 Santa 
Barbara St., Pasadena, CA 91101 {\&} Princeton University Observatory, 
Peyton Hall, Princeton, NJ 08544; \underline {iii@ociw.edu}

\newpage 
%\section{ABSTRACT}
%\label{sec:abstract}
\begin{center}
ABSTRACT
\end{center}
Recent radiative lifetime measurements accurate to $\pm $5{\%} (Stockett et 
al. 2007, J. Phys. B 40, 4529) using laser-induced fluorescence (LIF) on 8 
even-parity and 62 odd-parity levels of Er \textsc{ii} have been combined 
with new branching fractions measured using a Fourier transform spectrometer 
(FTS) to determine transition probabilities for 418 lines of Er \textsc{ii}. 
This work moves Er \textsc{ii} onto the growing list of rare earth spectra 
with extensive and accurate modern transition probability measurements using 
LIF plus FTS data. This improved laboratory data set has been used to 
determine a new solar photospheric Er abundance, log $\varepsilon $ = 0.96 
$\pm $ 0.03 ($\sigma $ = 0.06 from 8 lines), a value in excellent agreement 
with the recommended meteoric abundance, log $\varepsilon $ = 0.95 $\pm $ 
0.03. Revised Er abundances have also been derived for the $r-$process-rich 
metal-poor giant stars CS 22892-052, BD+17$^{o}$3248, HD 221170, HD 115444, 
and CS 31082-001. For these five stars the average Er/Eu abundance ratio, 
$<$log $\varepsilon $ (Er/Eu)$>$ = 0.42, is in very good agreement with the 
solar-system $r-$process ratio. This study has further strengthened the finding 
that $r-$process nucleosynthesis in the early Galaxy which enriched these 
metal-poor stars yielded a very similar pattern to the $r-$process which 
enriched later stars including the Sun. 

\smallskip
Subject headings: atomic data --- stars: abundances stars: Population II --- 
Sun: abundances--- Galaxy: evolution---nuclear reactions, nucleosynthesis, 
abundances---stars: individual (HD 115444, HD 221170, BD +17 3248, CS 
22892-052, CS 31082-001, CS 29497-030)

\newpage 
%\section{INTRODUCTION}
\begin{center}
1. INTRODUCTION
\end{center}
The study of elemental abundances in stellar photospheres continues to be a 
rich area of investigation. The availability of new large aperture 
telescopes has dramatically increased the number of target stars for which 
high spectral resolution data with a high signal-to-noise ratio can be 
obtained. One of the major successes in this area during recent years was 
the discovery and detailed study of a class of metal-poor Galactic halo 
stars with variable $n$(eutron)-capture elemental abundances (e.g. Sneden et 
al. 1995, Smith et al. 1995, Cowan et al. 1996, Sneden et al. 1996, Woolf et 
al. 1995, Sneden et al. 2000, Burris et al. 2000). Halo stars are among the 
oldest objects in the Galaxy, and provide a window on the earliest phases of 
Galactic evolution. The last decade has seen the first detection and 
abundance determination of numerous heavy elements in very metal-poor, 
$n$-capture rich halo stars, including the important chronometer uranium 
(Cayrel et al. 2001, Frebel et al. 2007).

Rare Earth (RE) elements are among the most spectroscopically accessible of 
the $n$-capture elements. The open f-shell of the RE neutral atoms and ions 
yields many strong lines in the visible and near-IR where spectral line 
blending is less of a problem than in the UV. Advantages from reduced 
blending in astrophysical data analysis are not matched by ease in 
calculating the basic atomic data needed for abundance determinations. These 
species with open f-shells have substantial relativistic effects causing a 
nearly complete breakdown of Russell-Saunders coupling\footnote{ 
Russell-Saunders coupling applies for light atoms in which the Coulomb 
repulsion of electrons in the Hamiltonian overwhelms relativistic effects 
including spin-orbit, spin-other orbit, spin-spin, and orbit-orbit 
interactions. In most levels of light atoms the total electronic angular 
momentum operator \textbf{L}$^{2}$ and total electronic spin 
\textbf{S}$^{2}$ are diagonal, or yield very good quantum numbers. In RE 
atoms relativistic effects are typically comparable to, or larger than, 
Coulomb repulsion terms in the Hamiltonian. The only generally good quantum 
numbers for levels of light atoms and RE atoms are eigenvalues of the total 
electronic angular momentum \textbf{J}$^{2}$ = (\textbf{L} + 
\textbf{S)}$^{2}$ and parity operators. }. They also have many low-lying, 
overlapping configurations leading to extensive configuration interaction. 
In some cases there are hundreds to thousands of interacting levels that 
need to be included in accurate calculations on the strongest 
``resonance-like'' transitions. Ab-initio quantum mechanical calculations on 
these spectra represent a formidable task even with the best currently 
available computers. The challenge of calculating spectroscopic data for RE 
neutral atoms and ions has attracted the attention of theorists (see 
Bi\'{e}mont {\&} Quinet 2003 and references therein). In such complex 
spectra progress is being made through an interplay of theory and 
experiment. Often some experimental information is essential to ``tune'' 
theoretical methods. 

The systematic determination of experimental transition probabilities by 
combining radiative lifetimes from time-resolved laser induced fluorescence 
(TR-LIF) with branching fractions from emission data recorded with a Fourier 
transform spectrometer (FTS) has played a central role in providing the 
basic atomic data needed for RE abundance determinations (e.g. Lawler 2006 
and references therein). This method yields absolute transition 
probabilities which are accurate to $\pm $5{\%} ($\sim $0.02 dex) for strong 
lines. Improved laboratory data has reduced line-to-line and star-to-star 
scatter in abundance values for many RE elements. The emergence of a tightly 
defined $r$-process only abundance pattern in many very metal-poor Galactic 
halo stars, at least for the RE elements, has been an exciting development 
(e.g. Sneden et al. 2003, Ivans et al. 2006, Lawler et al. 2006, Den Hartog 
et al. 2006). As this abundance pattern becomes even more tightly defined, 
it will: i) provide a powerful constraint on future modeling of the 
$r-$process nucleosynthesis; ii) help determine a definitive $r-$process site; and 
iii) unlock other details of the $r-$process and of the Galactic chemical 
evolution.

Erbium is one of the RE elements in need of additional work. There have been 
some LIF lifetime measurements (e.g. Bentzen et al. 1982, Xu et al. 2003, Xu 
et al. 2004), but a large set of experimental transition probabilities based 
on the best modern methods was not available before this work. Recent and 
extensive TR-LIF lifetime measurements by Stockett et al. (2007) provide a 
foundation for determining a large set of atomic transition probabilities 
from FTS data. We report the measurement of branching fractions for 418 
lines of Er II and the determination of absolute transition probabilities 
for these lines by combining our branching fractions with radiative lifetime 
data from Stockett et al. (2007). These laboratory data are applied to 
re-determine the Solar abundance of Er and to refine the Er abundance in 
five $r-$process rich, metal poor Galactic halo stars.

\begin{center}
2. Er \textsc{ii} BRANCHING FRACTIONS AND ATOMIC TRANSITION PROBABILITIES
\end{center}

The availability of large and accurate set of radiative lifetimes from 
Stockett et al. (2007) provides a foundation for this study of branching 
fractions and the transition probabilities of Er \textsc{ii}. A very 
powerful spectrometer is essential for branching fraction measurements on 
rich RE spectra. As in earlier work on RE spectra, we used the 1.0 meter FTS 
at the National Solar Observatory (NSO) for branching fraction measurements 
in this project. This instrument has the large etendue of all 
interferometric spectrometers, a limit of resolution as small as 0.01 
cm$^{-1}$, wavenumber accuracy to 1 part in 10$^{8}$, broad spectral 
coverage from the UV to IR, and the capability of recording a million point 
spectrum in 10 minutes (Brault 1976). An FTS is insensitive to any small 
drift in source intensity since an interferogram is a simultaneous 
measurement of all spectral lines.

2.1 Energy Levels of Er \textsc{ii} 

One of the challenges in this undertaking is the lack of configuration and 
term assignments for most observed levels of Er \textsc{ii}. Figure 1 shows 
a partial Grotrian diagram constructed from the compilation of Martin et al. 
(1978) for this ion. A total of 117 even-parity and 243 odd-parity levels 
are included in the compilation. The substantial overlap of low 
configurations leads to extensive configuration interaction and makes 
definitive assignments quite difficult for many levels. The lack of level 
assignments causes only minor difficulties in experimental work on branching 
fractions and transition probabilities since one cannot guess the strongest 
branches from an upper level. However, the lack of level assignments makes 
ab-initio theoretical determination of transition probability data very 
difficult.

Configuration and term assignments are firm for the lowest 26 levels of the 
117 known even-parity levels including: 12 levels of the 
4f$^{12}(^{3}$H)6s$_{1/2 }$and 4f$^{12}(^{3}$F)6s$_{1/2 }$ 
sub-configurations, 10 levels of the 4f$^{12}(^{3}$H$_{6})$5d$_{3/2 }$and 
4f$^{12}(^{3}$H$_{6})$5d$_{5/2 }$ sub-configurations, and 4 levels 
4f$^{12}(^{3}$F$_{4})$5d$_{3/2 }$ sub-configurations. Fortunately this 
list of low even-parity levels with firm assignments is nearly complete 
below $\sim $20,000 cm$^{-1}$. Although there is a missing 
4f$^{12}(^{1}$G)6s$_{1/2 }^{2}$G term in the 15,000 to 20,000 cm$^{-1}$ 
range, the nearly complete list of even-parity levels $<$ 20,000 cm$^{-1}$ 
reduces concerns of possible strong branches to unobserved low even-parity 
levels affecting the accuracy of our branching fraction measurements from 
upper odd-parity levels. Above 20,000 cm$^{-1}$ there are numerous 
unobserved even-parity levels. Between 25,000 cm$^{-1}$ and 31,000 cm$^{-1}$ 
there are only 9 levels assigned to the 4f$^{12}(^{3}$H)5d$_{ }$ 
sub-configuration and these lack term assignments. Except for the 9 levels 
of the 4f$^{11}(^{3}$H$_{15/2})$6s6p($^{3}$P) sub-configuration between 
32,000 cm$^{-1}$ and 38,000 cm$^{-1}$ the remaining even-parity levels are 
either tentatively assigned to the 4f$^{11}(^{4}$I)5d6p 
sub-configuration$_{ }$ or in most cases unassigned. A new analysis of Er 
\textsc{ii} is underway (Wyart et al. 2008 to be submitted).

The fraction of observed levels with assignments is somewhat lower for the 
243 known odd-parity levels. Only the lowest odd-parity level at 6825 
cm$^{-1}$ and two higher levels have firm term assignments. These three 
levels are part of the low $^{4}$I term of 4f$^{11}$6s$^{2}$ configuration. 
Another 28 levels starting from 10,667 cm$^{-1}$ have firm assignments to 
the 4f$^{11}(^{4}$I)5d6s sub-configuration and seven have tentative 
assignments to this sub-configuration. These 38 levels are the lower levels 
of strong branches from the upper even-parity levels included in our 
branching fraction study. There are 10 additional levels ranging from 25,000 
cm$^{-1}$ to 34,000 cm$^{-1}$ with tentative term and configuration 
assignments to the 4f$^{12}$6p configuration. All other odd-parity levels 
lack both term and configuration assignments. The ongoing reanalysis of Er 
\textsc{ii} indicates the lowest unobserved odd-parity level is just under 
20,000 cm$^{-1}$ (Wyart et al. 2008 to be submitted). There are quite a 
number of unobserved odd-parity levels in the 20,000 cm$^{-1}$ to 30,000 
cm$^{-1}$ range. These levels, like many of the upper odd-parity levels 
included in our branching fraction study, are mixtures of states from 
4f$^{12}$6p, 4f$^{11}$6s$^{2}$, 4f$^{11}$5d6s, and 4f$^{11}$5d$^{2}$ 
configurations.

The crucial issue in this review of assignments for low Er \textsc{ii} 
levels is whether or not there are significant branches from upper levels in 
this study to unobserved lower levels. Although many previously unobserved 
levels have been located in the ongoing reanalysis of Er \textsc{ii }(Wyart 
et al. 2008 to be submitted), these levels are very weakly connected to 
upper levels of this study with one exception. We will return to the issue 
of unobserved levels after discussing our branching fraction measurements.

2.2 Er \textsc{ii} Branching Fraction Analysis and Relative Radiometric 
Calibration

As in earlier studies our experimental branching fractions are based on a 
large set of FTS data including: spectra of lamps at high currents to reveal 
very weak branches to known levels, good IR spectra to reveal any 
significant IR branches to known levels, and low current spectra in which 
dominant branches are optically thin covering the UV to near-IR. Table 1 is 
a list of the 15 FTS spectra used in our branching fraction study. All were 
recorded using the National Solar Observatory 1.0 meter FTS on Kitt Peak. 
Some of these spectra ({\#}1-6, 12-15) were recorded by other guest 
observers in the 1980's, and others ({\#}7-11) were recorded during our 
February 2000 and February 2002 observing runs. All 15 raw spectra are 
available from the electronic archives of the National Solar 
Observatory\footnote{Available at http://nsokp.nso.edu/}.

The establishment of an accurate relative radiometric calibration or 
efficiency is critical to a branching fraction experiment. As indicated in 
Table 1, we made use of both standard lamp calibrations and Ar \textsc{i} 
and Ar \textsc{ii} line calibrations in this Er \textsc{ii} study. Tungsten 
(W) filament standard lamps are particularly useful near the Si detector 
cutoff in the 10,000 to 9,000 cm$^{-1}$ range where the FTS sensitivity is 
changing rapidly as a function of wave number, and near the dip in 
sensitivity at 12,500 cm$^{-1}$ from the aluminum coated optics. Tungsten 
lamps are not bright enough to be useful for FTS calibrations in the UV 
region, and UV branches typically dominate the decay of levels studied using 
our lifetime experiment. In general one must be careful when using continuum 
lamps to calibrate the FTS over wide spectral ranges, because the ``ghost'' 
of a continuum is a continuum. The Ar \textsc{i} and Ar \textsc{ii} line 
technique, which is internal to the Hollow Cathode Discharge (HCD) Er/Ar 
lamp spectra, is still our preferred calibration technique. It captures the 
wavelength-dependent response of detectors, spectrometer optics, lamp 
windows, and any other components in the light path or any reflections which 
contribute to the detected signal (such as due to light reflecting off the 
back of the hollow cathode). This calibration technique is based on a 
comparison of well-known branching ratios for sets of Ar \textsc{i} and Ar 
\textsc{ii} lines widely separated in wavelength, to the intensities 
measured for the same lines. Sets of Ar \textsc{i} and Ar \textsc{ii} lines 
have been established for this purpose in the range of 4300 to 35000 
cm$^{-1}$ by Adams {\&} Whaling (1981), Danzmann {\&} Kock (1982), 
Hashiguchi {\&} Hasikuni (1985), and Whaling et al. (1993). One of our best 
Er/Ar HCD spectra from 2002, and the Er/Ar HCD spectra from 1987 and 1988, 
were calibrated with both W standard lamp spectra recorded shortly before, 
or after, the HCD lamp spectra and using the Ar\textsc{ i} and Ar 
\textsc{ii} line technique. The Er/Ne spectra from the 1987 and 1988 could 
only be calibrated using W standard lamp. The older W lamp is a strip lamp 
calibrated as a spectral radiance (W/(m$^{2}$ sr nm)) standard, and the 
newer is a tungsten-quartz-halogen lamp calibrated as a spectral irradiance 
(W/(m$^{2}$ nm) at a specified distance) standard. Neither of these W 
filament lamps is hot or bright enough to yield a reliable UV calibration, 
but they are useful in the visible and near IR for interpolation and as a 
redundant calibration.

All possible transition wave numbers between known energy levels of Er 
\textsc{ii} satisfying both the parity change and $\Delta $J = -1, 0, or 1 
selection rules were computed and used during analysis of FTS data. Energy 
levels from Martin et al. (1978) were used to determine possible transition 
wave numbers. Levels from Martin et al. (1978) are available in electronic 
form from Martin et al. (2000)\footnote{Available at http:// 
physics.nist.gov/PhysRefData/ASD/index.html}. Systematic errors from missing 
branches to known lower levels are negligible in our work, because we were 
able to make at least rough measurements on ultraviolet through IR lines 
with branching fractions of 0.001 or smaller. This is illustrated in Table 2 
which lists our branching fractions for the odd-parity upper level at 
28361.386 cm$^{-1}$. For this level we were able to measure and report a 
very weak, 0.00040, branching fraction. Figures 2 and 3 show some Er 
\textsc{ii} line profiles from this upper level. Figure 2 is the shortest 
wavelength, second strongest transition at 3524.905 {\AA} (branching 
fraction 0.389). Figure 3 is the second longest wavelength transition at 
15458.105 {\AA} (branching fraction 0.00122). Given the large wavelength 
separation of these two lines, it should not be surprising that the data in 
Figures 2 and 3 are from different spectra. Isotopic structure is clearly 
visible in the IR line of Figure 3. The triplet pattern is due to the even 
(nuclear spin I = 0) isotopes $^{166}$Er (abundance 33.61{\%}), $^{168}$Er 
(abundance 26.78{\%}), and $^{170}$Er (abundance 14.93{\%}) (Rossman {\&} 
Taylor 1997). Hyperfine structure ``smears out'' the transition of the other 
abundant Er isotope which is the odd (I = 7/2) isotope $^{167}$Er (abundance 
22.93{\%}) and individual hyperfine components from this odd isotope are 
difficult to detect in our spectra. The two lightest isotopes, $^{164}$Er 
(abundance 1.61{\%}) and $^{162}$Er (abundance 0.14{\%}), have such low 
abundances that they are not detectable in our spectra. Isotopic splittings 
are somewhat larger in the IR than in the UV for lines studied, but one 
should keep in mind that the FTS data of Figure 3 has higher spectral 
resolution than that of Figure 2. The IR lines with relatively large isotope 
shifts are very weak and have such large excitation potentials that we see 
no hope of detecting the lines in astrophysical spectra for the foreseeable 
future.

Branching fraction measurements were attempted on lines from all 80 levels 
of the lifetime experiment by Stockett et al. (2007), and were completed for 
lines from 7 even-parity and 63 odd-parity upper levels. The levels for 
which branching fractions could not be completed had a strong branch beyond 
the UV limit of our spectra, or had a strong branch which was severely 
blended. Typically an odd-parity upper level, depending on its J value, has 
about 20 possible transitions to known lower levels, and an even-parity 
upper level has about 60 possible transitions to known lower levels. More 
than 20,000 possible spectral line observations were studied during the 
analysis of 15 different Er/Ar and Er/Ne spectra. We set integration limits 
and occasionally nonzero baselines ``interactively'' during analysis of the 
FTS spectra. An occasional nonzero baseline is needed when a weak line is 
located on a line wing of a much stronger line. The same numerical 
integration routine was used to determine the un-calibrated intensities of 
Er \textsc{ii} lines and selected Ar \textsc{i} and Ar \textsc{ii} lines 
used to establish a relative radiometric calibration of the spectra. A 
simple numerical integration technique was used in this and most of our 
other RE studies because of weakly resolved or unresolved hyperfine and 
isotopic structure. More sophisticated profile fitting is used only when the 
line sub-component structure is either fully resolved in the FTS data or 
known from independent measurements.

2.3 Branching Fraction Uncertainties

The procedure for determining branching fraction uncertainties was described 
in detail by Wickliffe et al. (2000). Branching fractions from a given upper 
level are defined to sum to unity, thus a dominant line from an upper level 
has small branching fraction uncertainty almost by definition. Branching 
fractions for weaker lines near the dominant line(s) tend to have 
uncertainties limited by their S/N ratios. Systematic uncertainties in the 
radiometric calibration are typically the most serious source of uncertainty 
for widely separated lines from a common upper level. We used a formula for 
estimating this systematic uncertainty that was presented and tested 
extensively by Wickliffe et al. (2000). The spectra of the high current 
custom HCD lamps enabled us to connect the stronger visible and near IR 
branches to quite weak branches in the same spectral range. Uncertainties 
grew to some extent from piecing together branching ratios from so many 
spectra, but such effects have been included in the uncertainties on 
branching fractions of the weak visible and near IR lines. In the final 
analysis, the branching fraction uncertainties are primarily systematic. 
Redundant measurements with independent radiometric calibrations help in the 
assessment of systematic uncertainties. Redundant measurements from spectra 
with different discharge conditions also make it easier to spot blended 
lines and optically thick lines. Many of the strong lines in the UV and 
visible were optically thick in the spectra from the Custom HCD lamp 
operating at high current. These data were discarded during review of the 
branching ratio data before combining data from the various spectra to 
determine our final branching fractions.

As mentioned in {\S}2.1, one of the more troubling systematic uncertainties 
is from possible branches to unobserved lower levels. We have checked for 
branches from upper levels in this study to previously unobserved lower 
levels using both an experimental search to tentatively identified lower 
levels, and using results from a parametric fit to the energy levels. With 
only one exception, the upper levels of this study are very weakly connected 
to the unobserved lower levels. Based on the reanalysis of Er \textsc{ii} to 
date, only the highest upper level of this study, the even-parity level at 
46757.780 cm$^{-1}$, is likely to have significant branches to unobserved 
odd-parity lower levels (Wyart et al. 2008 to be submitted). Transition 
probabilities from this upper level have been reduced by 7.7{\%} ($\sim 
$0.03 dex) to correct for the branches to unobserved lower levels. This 
correction introduces some additional systematic uncertainty for the four 
lines from this upper level included in our study. The reanalysis indicates 
that the odd-parity upper level at 33307.365 cm$^{-1}$ has J = 3.5 (7/2 in 
standard notation) instead of 4.5 (9/2) as given in the NIST tables (Martin 
et al. 1978)\footnote{ Redundant decimal notation and standard fractional 
notation for J values are included in the text, but our tables use only 
decimal notation required for the main machine readable table of transition 
probabilities. }. The Land\'{e} g-factor supports this change. Careful 
inspection of all spectra in this study revealed some weak lines from this 
upper level to J = 2.5 (5/2) lower level and not a hint of a transition to a 
lower level with J = 5.5 (11/2). We therefore use the modified J = 3.5 (7/2) 
for the level at 33307.365 cm$^{-1}$ and note that this change does not 
affect our Einstein A-coefficients from this upper level, but does affect 
the log(\textit{gf}) values from this upper level. The reanalysis also indicates that 
the J = 4.5 (9/2) odd-parity level at 33129.912 cm$^{-1}$ is not real. Only 
a single emission line at 3570.75 {\AA} from this upper level was detected 
in our branching fraction study, and this level does not fit in the 
parametric study of Er \textsc{ii} (Wyart et al. to be submitted). The 
lifetime of 4.7 ns reported by Stockett et al. (2007) is correct for laser 
excitation at 3570.75 {\AA}, but no transition probabilities can be reported 
until the J and actual energy of the upper level is established.

2.4 Er \textsc{ii} Atomic Transition Probabilities

Branching fractions from the FTS spectra were combined with the radiative 
lifetime measurements (Stockett et al.2007) to determine absolute transition 
probabilities for 418 lines of Er \textsc{ii} in Table 3. Air wavelengths in 
Table 3 were computed from energy levels (Martin et al. 1978) using the 
standard index of air (Edl\'{e}n 1953). Parities are included in Table 3 
using ``ev'' and ``od'' notation which is compatible with our main machine 
table of transition probabilities. 

Transition probabilities for the very weakest lines (branching fractions 
$\sim $ 0.001 or weaker) which were observed with poor S/N ratios and for a 
few blended lines are not included in Table 3, however these lines are 
included in the branching fraction normalization. The effect of the problem 
lines becomes apparent if one sums all transition probabilities in Table 3 
from a chosen upper level, and compares the sum to the inverse of the upper 
level lifetime from Stockett et al. (2007). Typically the sum of the Table 3 
transition probabilities is between 95{\%} and 100 {\%} of the inverse 
lifetime. Although there is significant fractional uncertainty in the 
branching fractions for these problem lines, this does not have much effect 
on the uncertainty of the stronger lines that were kept in Table 3. 
Branching fraction uncertainties are combined in quadrature with lifetime 
uncertainties to determine the transition probability uncertainties in Table 
3. 

There are only a few comparisons which can be made between our transition 
probability data and other similar data. The most interesting comparison is 
to the experimental work of Musiol and Labuz (1983) shown in Figure 4. The 
discordant points of Figure 4 may be, in some cases, due to incorrect line 
identifications from the lower resolving power achieved in the earlier 
grating spectrometer measurements by Musiol and Labuz. In complex rare earth 
spectra, the resolution and absolute wavenumber accuracy of a FTS is 
extremely important. Line broadening and blending could also have been a 
problem in the experiments by Musiol and Labuz because they used a high 
pressure (LTE) arc plasma. Lines from our low pressure HCD lamps are 
primarily Doppler broadened in most cases. Although the comparison to Musiol 
and Labuz is not as favorable as one might hope, it is better than the 
comparisons to theoretical results in Figures 5 and 6. Figures 5 and 6 are, 
respectively, comparisons of our results against relativistic Hartree Fock 
calculations (Xu et al. 2003) and semi-empirical results from Kurucz (2007). 
It is important to recall that the very comprehensive Kurucz database was 
originally intended for opacity calculations, and not for precise 
spectroscopic research. (It should also be noted that some of the Kurucz 
data is from Labuz and Musiol.) Calculations of transition probabilities in 
Er \textsc{ii} are indeed a very difficult theoretical undertaking. We note 
that the reanalysis of Er \textsc{ii} is yielding encouraging results. 
Theoretical branching fractions are in good agreement with experimental 
branching fractions for all of the even-parity upper levels in this study, 
and for about half of the odd-parity upper levels. In the next sections we 
apply our new laboratory results in Er abundance determinations. 

\begin{center}
3. SOLAR AND STELLAR ERBIUM ABUNDANCES
\end{center}

The new transition probabilities have been applied to Er \textsc{ii} lines 
in the solar photosphere and five very metal-poor ([Fe/H] $<$ -2)\footnote{ 
We adopt standard stellar spectroscopic notations that for elements A and B, 
[A/B] = log$_{10}$(N$_{A}$/N$_{B})_{star}$ - 
log$_{10}$(N$_{A}$/N$_{B})$sun, for abundances relative to solar, and log 
$\varepsilon $(A) = log$_{10}$(N$_{A}$/N$_{H})$ + 12.0, for absolute 
abundances.} stars that have large overabundances of the rare earth 
elements. Our abundance study followed the methods used for Hf \textsc{ii} 
by Lawler et al. (2007) and previous papers in this series. Erbium has been 
less well studied in solar/stellar spectra than have many other rare earth 
ions, due to a lack of extensive previous lab investigations and to a 
paucity of transitions in spectral regions convenient for ground-based high 
resolution spectroscopy. Anecdotal evidence to support this suggestion comes 
from the classic Moore, Minnaert, {\&} Houtgast (1966) solar line 
compendium. Those authors could identify only 2 Er \textsc{ii} transitions 
(at 3896.2 {\AA} and 3903.3 {\AA}), in contrast with the large number they 
identified for many other rare-earth ions (e.g. 146 Sm \textsc{ii} and 72 Gd 
\textsc{ii} lines). Identification of a suitable set of Er \textsc{ii} lines 
was therefore as important as the subsequent abundance analysis. 

3.1 Line Selection

We have accurate transition probabilities for 418 Er \textsc{ii} lines, but 
only a small minority of these can be employed to determine Er abundances in 
the Sun and our chosen metal-poor stars. This is because all strong Er 
\textsc{ii} lines occur only in the near-UV spectral domain, $\lambda  \quad <$ 
4000 {\AA}. As discussed by Lawler et al. (2007 and references therein), to 
first approximation the relative strengths of weak-to-moderate lines within 
one species depend directly on their transition probabilities modified by 
the Boltzmann excitation factors. For a line on the linear part of the 
curve-of-growth the relationship between equivalent width EW, reduced width 
RW, transition probability, excitation energy $\chi $ (measured in eV), and 
inverse temperature $\theta  \quad \equiv $ 5040/T is: 

log(RW) = log(EW/$\lambda )$ = constant + log(\textit{gf}) -- $\theta \chi {\rm g}$ 

The relative strengths of lines of different species also depend on relative 
elemental abundances and Saha ionization equilibrium factors. However, the 
relatively low first ionization potential of Er (6.108 eV, Grigoriev {\&} 
Melikhov 1997) ensures that it almost entirely exists as Er \textsc{ii} in 
the photospheres of the Sun and stars considered here. Therefore Er 
\textsc{ii}, like those of all rare-earth single ions, needs essentially no 
Saha corrections for the existence of other ionization states. Thus for all 
elements with similarly low ionization potentials their weak-ionized-line 
strength factors are 

STR $\equiv $ log($\varepsilon $\textit{gf}) $-- \theta \chi $,  

where $\varepsilon $ is the elemental abundance.

In Figure 7 we plot these relative strength factors as a function of 
wavelength for Gd \textsc{ii} lines (Den Hartog et al. 2006) and Er 
\textsc{ii} lines (this paper). To compute the strength factors we have 
adopted solar abundances of log $\varepsilon $(Gd) = +1.11 (Den Hartog et 
al.) and log $\varepsilon $(Er) = +0.95 (close to the recommended 
photospheric abundance of Grevesse {\&} Sauval 2002 and Lodders 2003) which 
will be the new value derived in this paper. This plot is very similar to 
ones that we have shown in several of our previous papers. As in those 
studies we have used horizontal lines to indicate approximate strength 
factors for ``strong'' and ``barely detectable'' lines as follows.

The minimum detectable strength limit for Sm \textsc{ii} lines was estimated 
by Lawler et al. (2006) by first searching the Delbouille et al. (1973) 
solar photospheric spectrum for the weakest lines that could be reliably 
employed in an abundance analysis. That exercise suggested an EW limit of 
about 1.5 m{\AA} near $\lambda  \quad \sim $ 4500 {\AA}, or log(RW) $\approx $ 
-6.5. Lines of Sm \textsc{ii} near this limit had have STR = 
log($\varepsilon $ \textit{gf}) -- $\theta \chi  \quad \approx $ -0.6 . That EW and thus 
STR limit should apply also to Gd \textsc{ii} and Er \textsc{ii} lines, and 
so it has been indicated in both panels of Figure 7 with horizontal dotted 
lines. 

Minimum strength factors for relatively strong lines were estimated by 
Lawler et al. (2006) by beginning with the detection-limit STR = -0.6 and 
increasing it by a factor of 20, or STR = -0.6 + 1.3 = +0.7. Ignoring 
curve-of-growth saturation effects would imply that log(RW) = -6.5 + 1.3 = 
-5.2 (or EW $\approx $ 30 m{\AA} near 4500 {\AA}). Such lines actually are 
slightly saturated, and tests with the solar spectrum suggested log(RW) 
$\approx $ -5.35, or EW $\approx $ 20 m{\AA} at 4500 {\AA} for STR = +0.7. 
We have drawn dashed horizontal lines to indicate this ``strong-line'' limit 
in Figure 7.

This study has reported transition probabilities of Er \textsc{ii} lines 
with wavelengths nearly as long as 20,000 {\AA} (2 $\mu )$, but Figure 7 
displays only the regime 2900 {\AA} $\le  \quad \lambda  \quad \le $ 6000 {\AA}. 
This is because all of the Er \textsc{ii} lines beyond 6000 {\AA} have STR 
$<$ -1.7, more than 1 dex weaker than our estimated minimum detectability 
threshold of STR = -0.6. In fact, the right-hand panel of this figure shows 
that very few Er \textsc{ii} lines should even be detectable in the solar 
spectrum longward of 4000 {\AA}. We have drawn vertical lines at 4000 {\AA} 
in the figure to bring attention to this difficulty. Nearly 75 Gd 
\textsc{ii} lines longward of 4000 {\AA} have STR $\ge $ -0.6, while just 6 
Er \textsc{ii} lines qualify. All strong Er \textsc{ii} lines are located in 
the complex near-UV spectral domain, where line blending from other species 
might compromise even the most promising Er \textsc{ii} transition.

As discussed in Lawler et al. (2007) and earlier papers, the strength 
factors of Figure 7 provided the first cut in paring the list of 418 Er 
\textsc{ii} lines to a useful set for solar/stellar work. Some 115 lines 
survived the STR $\ge $ -0.6 test. We then followed Lawler et al. to 
identify the final set of potentially useful Er \textsc{ii} lines. Using the 
Delbouille et al. (1973) solar center-of-disk spectrum, the Moore et al. 
(1966) solar line identifications, the comprehensive Kurucz 
(1998)\footnote{Available at http://kurucz.harvard.edu/} atomic and 
molecular line lists, and the spectrum of the $r-$process-rich metal-poor giant 
star CS 22892-052 (Sneden et al. 2003), we eliminated all but 57 Er 
\textsc{ii} lines; the rest proved to be too weak and/or too blended (see 
Lawler et al. 2006 for specific examples of the process). The CS 22892-052 
spectrum was especially helpful in this exercise, as the combined effects of 
its very low metallicity ([Fe/H] $\approx $ -3.1) and large neutron-capture 
$r$-process excess (e.g., [Eu/Fe] $\approx $ +1.6) creates very favorable 
conditions for Er \textsc{ii} line detection. If a candidate line is 
unusable in the CS 22892-052 spectrum, it almost certainly will not be 
available for a solar analysis.

We then computed preliminary synthetic spectra for the surviving Er 
\textsc{ii} lines. As in Lawler et al. (2006), we assembled atomic and 
molecular line lists in small (4-6 {\AA}) wavelength regions, beginning with 
Kurucz's (1998) line database and Moore et al.'s (1966) solar 
identifications. For many neutron-capture ionized species we used 
\textit{gf}-values from recently published studies: Y, Hannaford et al. (1982); Zr, 
Malcheva et al. (2006); La, Lawler et al. (2001a); Ce, Palmeri et al. 
(2000); Nd, Den Hartog et al. (2003); Sm, Lawler et al. (2006); Eu, Lawler 
et al. (2001b); Gd, Den Hartog et al. (2006); Tb, Lawler et al. (2001c); Dy, 
Wickliffe et al. (2000); Ho, Lawler et al. (2004); Er, the present paper; 
and Hf, Lawler et al. (2007). We adopted the Holweger {\&} M\"{u}ller (1974) 
solar empirical model photosphere, and the CS 22892-052 model interpolated 
from the Kurucz grid by Sneden et al. (2003). For solar computations we used 
a standard solar abundance set (e.g. Grevesse {\&} Sauval 1998, 2002; 
Lodders 2003), modified to include recent updates for the neutron-capture 
elements, and for CS 22892-052 we used abundances from Sneden et al. (2003), 
modified for neutron-capture elements by our previous papers in this series.

Line lists, model atmospheres, and abundance sets were input into the 
current version of the LTE line analysis code MOOG (Sneden 1973) to generate 
initial synthetic spectra. Empirical Gaussian broadening functions were 
applied to smooth the synthetic spectra to match the effects of 
solar/stellar macroturbulence and spectrograph instrumental profile. Visual 
inspection of the synthetic/observed spectrum matches were sufficient to 
reduce the 57 candidate lines to 23 that were suitable for abundance 
analysis in the Sun and/or CS 22892-052. These transitions were the ones 
examined in all program stars.

3.2 The Solar Photospheric Erbium Abundance

We computed multiple synthetic spectra for each Er line region in a more 
careful manner, trying to account for the details of the solar spectra. As 
discussued in {\S}2, Er \textsc{ii} lines in the red-IR have detectable 
hyperfine/isotopic substructure (Figure 3), but it is negligible for the 
near-UV lines (e.g., Figure 2) that we used for solar/stellar abundances. 
Therefore we treated these lines as single absorbers. The oscillator 
strengths for atomic lines other than the neutron-capture species referenced 
in {\S}3.1 were adjusted to fit the solar spectrum. Abundances of elements 
C, N, and O were altered to match the strengths of observed CH, CN, NH, and 
OH lines. Of course many solar absorption features, especially in the 
near-UV spectral region of greatest interest in this study, remain 
unidentified. We arbitrarily declared these lines to be Fe \textsc{i} with 
excitation potentials $\chi $ = 3.5 eV and \textit{gf}-values adjusted to fit the solar 
spectrum. We compared these iterated synthetic spectra to the Delbouille et 
al. (1973) center-of-disk photospheric spectrum. In any case where line 
contamination of identified or unknown origin proved to be a significant 
part of the overall absorption at the Er \textsc{ii} wavelength, the line 
was discarded for the solar analysis but kept for possible use with the 
metal-poor giants.

The final solar Er abundance is based on eight Er \textsc{ii} lines, whose 
individual abundances are listed in Table 4, column 4. These lines include 
the 3896.2 {\AA} feature identified by Moore et al. (1966), but their 3903.3 
{\AA} line was not part of our laboratory investigation. In the top panel of 
Figure 8 we display the solar Er line abundances; no obvious trends with 
wavelength are apparent. A straight mean abundance is log $\varepsilon $(Er) 
= 0.96 $\pm $ 0.02 ($\sigma $ = 0.06).

Abundance uncertainties have been described in earlier papers of this 
series. Here, we estimate line profile fitting uncertainties to be $\pm 
$0.02 dex, and uncertainties due to contamination by other species lines are 
$\pm $0.04 dex. The mean error in log(\textit{gf}) for the eight lines used in the 
solar analysis (see Table 3) is $\pm $0.02. Adding these uncertainties in 
quadrature yields an estimated total internal uncertainty per transition of 
$\pm $0.05 dex, which is close to the observed $\sigma $ = 0.06. 

Overall scale errors can be due to atomic data uncertainties beyond \textit{gf} errors, 
and model atmosphere choices. Recalling that Saha-fraction corrections are 
negligible for Er \textsc{ii}, the main atomic uncertainties would be 
Boltzmann factors, which vary with the partition functions. Irwin (1981) 
computed polynomial fits to partition functions that were generated with the 
atomic energy level data available at that time, and his formulae have been 
widely used in stellar line analysis programs. We re-calculated Er 
\textsc{i} and Er \textsc{ii} partition functions with the most recent 
experimental energy level data (Martin et al. 1978, 2000), and found that 
the new values of log(U) are up to $\sim $0.2 dex larger in the temperature 
domain of interest for this study. We have used the new partition functions 
from experimental energy levels for all of our abundances. A reanalysis of 
Er \textsc{ii}, which is currently underway (Wyart et al. 2008 to be 
submitted), indicates that the remaining unobserved low-lying levels of each 
parity could further increase the partition function by 0.016 dex at 6000 K, 
but much less at lower temperature. This final theoretical correction to the 
Er \textsc{ii} partition function is not included here because nearly all of 
the rare earth partition functions need similar adjustments due to 
unobserved levels. 

As in Lawler et al. (2007), we repeated some of the abundance computations 
using the Kurucz (1998) and Grevesse {\&} Sauval (1999) models, finding on 
average abundance shifts of -0.02 dex compared to those done with the 
Holweger {\&} M\"{u}ller (1974) model. Combining line-to-line scatter 
uncertainties ($\pm $0.02 from the standard deviation of the mean, Table 4) 
with scale uncertainties, we recommend log $\varepsilon $(Er)$_{Sun}$ = 
+0.96 $\pm $ 0.03.

Bi\'{e}mont {\&} Youssef (1984) provided the previous major solar Er 
investigation. From an equivalent width analysis of seven lines , they 
derived log$\varepsilon $(Er)$_{Sun}$ = +0.93 $\pm $ 0.06, in good agreement 
with our new value (only the 3781.0 {\AA} and 3896.2 {\AA} lines are in 
common between the two studies). Lodders (2003) adopts this abundance in her 
solar abundance review, and recommends a meteoritic value in even closer 
agreement with our value: log$\varepsilon $(Er) = +0.95 $\pm $ 0.03. This 
point will be considered again in {\S}3.4. 

In Figure 9 we compare solar-system meteoritic abundances with photospheric

abundances for the nine rare-earth elements studied in this series of 
papers. The meteoritic values are adopted from Lodders' (2003) compilation. 
References to the photospheric values are given in the figure caption. It is 
clear that the two data sets agree well: a simple mean offset is +0.01 $\pm 
$ 0.02 (sigma = 0.05). These numbers are consistent with error estimates on 
individual meteoritic and photospheric abundances.

3.3 Erbium Abundances in Five $r-$Process-Rich Low Metallicity Stars

We also derived Er abundances in five very metal-poor, $r-$process-rich giant 
stars: CS 22892-052 ([Fe/H] = -3.1, [Eu/Fe] = +1.5, Sneden et al. 2003); 
BD+17$^{o}$3248 ([Fe/H] = -2.1, [Eu/Fe] = +0.9, Cowan et al. 2002); HD 
221170 ([Fe/H] = -2.2, [Eu/Fe] = +0.8, Ivans et al. 2006); and HD 115444 
([Fe/H] = -2.9, [Eu/Fe] = +0.8, Westin et al. 2000); CS 31082-001 ([Fe/H] = 
-2.9, [Eu/Fe] = +1.7, Hill et al. 2002). Many Er \textsc{ii} lines that are 
too blended and/or weak in the solar spectrum could be employed here, and we 
ended up with 14-21 lines contributing to the mean abundances. We derived Er 
abundances for the stars in the same manner as was described for the Sun in 
{\S}3.2. The abundances from individual lines are listed in Table 4 and 
displayed in Figure 8. The mean abundances, standard deviations, and number 
of lines are given at the bottom of Table 4 and Figure 8. The line-to-line 
scatters are all small, $\sigma $ = 0.04 -- 0.08. The derived Er abundances 
show no noticeable dependence on wavelength, log(\textit{gf),} or excitation potential.

\begin{center}
4. DISCUSSION
\end{center}

The Er results are similar to those found for other RE studies, where the 
new experimental atomic data has dramatically reduced the scatter in 
star-to-star elemental abundance comparisons. We show this agreement and 
these comparisons in Figure 10. In the top panel of the figure we show the 
differential elemental abundance values for the four stars CS 22892-052, 
BD+17 3248, HD 221170 and HD 115444. In all cases the stars' elemental 
abundances have been scaled relative to Eu and the differences are with 
respect to the predicted solar system $r-$process only value. For these cases we 
have employed the $r-$process predictions from Simmerer et al. (2004) (see also 
Sneden, Cowan, {\&} Gallino 2008). (A perfect agreement with the $r-$process 
only would fall on the dotted horizontal line in Figure 10.) Previous 
studies of the RE elements, including that of Nd (Den Hartog et al. 2003), 
Ho (Lawler et al. 2004), Sm (Lawler et al. 2006), and Gd (Den Hartog et al. 
2006) and of the inter-peak element Hf (Lawler et al. 2007), have improved 
the precision of the stellar elemental abundances, as is apparent by the 
close agreement in the figure. The Er abundances for these four studied 
stars now are tightly clustered with relatively small error bars indicated 
in the bottom of the figure (as means of the sigmas for each element in the 
four stars) and are consistent with the solar system $r-$process only value. 

Table 4 includes analyses for two additional stars. While not included in 
Figure 10 we have also analyzed the Er abundances in CS 31082-001. 
Previously we had determined the Hf abundance in this star (see Lawler et 
al. 2007). From 19 Er \textsc{ii} lines we derive log $\varepsilon $(Er) = 
-0.30 $\pm $ 0.01 ($\sigma $ = 0.04). With our own analysis of Eu 
\textsc{ii} lines (Lawler et al. 2007) we find log $\varepsilon $(Eu) = 
-0.72, and log $\varepsilon $(Eu/Er) = -0.42. This value is essentially 
identical to the Eu/Er ratios found for the four other $r-$process rich stars. A 
more complete analysis is underway (Ivans et al. to be submitted). In 
contrast to these $r-$process rich stars we have also measured nine Er\textsc{ 
ii} lines in the star CS 29497-030, a star rich in both $r-$ and $s-$ process 
material (Ivans et al. 2005). We find log $\varepsilon $(Er) = +0.57 $\pm $ 
0.02 ($\sigma $ = 0.07), or log $\varepsilon $(Eu/Er) = -0.64. The 0.2 dex 
difference in the ratio between this star and the other five stars results 
from the effects of changing from a pure $r$ abundance to a mix of $r+s$. 

Interestingly, it appears that the average Er abundance for the four 
$r-$process rich stars illustrated in Figure 10 lies just slightly above the 
scaled solar value. This might suggest that Er may be similar to the cases 
of Gd (Den Hartog et al. 2006) and Hf (Lawler et al. 2007), where the 
stellar data argue for a somewhat larger $r-$process fraction for the total 
solar system abundances (see also Sneden et al. 2008). Some of the remaining 
small systematic uncertainties, e.g. the correction of partition functions 
for unobserved levels, will further enhance the $r-$process abundances.

A few RE elements remain to be improved including Ce (Lawler et al., in 
preparation), but most of these elements have now been well studied. The Er 
results presented here, along with the other RE studies, have all led to 
much more precise stellar elemental abundances. These abundances in the 
metal-poor ($r-$process rich) halo stars are all consistent with a solar system 
$r-$process only origin. This study has further strengthened the finding that 
$r-$process nucleosynthesis in the early Galaxy which enriched these metal-poor 
stars yielded a very similar pattern to the $r-$process which enriched later 
stars including the Sun. This in turn provides important constraints on the 
timescales for such synthesis - that is it suggests rapidly evolving 
astronomical sites, forming the elements, ejecting them and mixing them into 
the interstellar medium, all prior to the formation of the halo stars - and 
points to the (possibly massive) nature of the first stars. 

\begin{center}
ACKNOWLEDGEMENTS
\end{center}

This work has been supported by the National Science Foundations through 
grants AST-0506324 to JEL {\&} EDH, AST- 0607708 to CS, and AST-0707447 to 
JJC.

\newpage 
\begin{center}
REFERENCES
\end{center}

Adams, D. L., {\&} Whaling, W. 1981, J. Opt. Soc. Am., 71, 1036

Bentzen, S. M., Nielsen, U., {\&} Poulsen, O. 1982, J. Opt. Soc. Am., 72, 
1210

Bi\'{e}mont, E. {\&} Youssef, N. Y. 1984, A{\&}A, 140, 177

Bi\'{e}mont, E. {\&} Quinet P. 2003, Physica Scripta, T105, 38

Brault, J. W. 1976, J. Opt. Soc. Am., 66, 1081

Burris D. L., Pilachowski C.A., Armandroff T. E., Sneden C., Cowan J. J., 
{\&} Roe H. 2000, ApJ, 544, 302 

Cayrel R., Hill V., Beers T. C., Barbuy B., Spite M., Spite F., Plez B., 
Andersen J., Bonifacio P., Francois P., Molaro P., Nordstrom B., {\&} Primas 
F. 2001, Nature 409, 691

Cowan J. J., Sneden C., Truran J. W., {\&} Burris D. L. 1996, ApJ, 460, L115 

Cowan, J. J., Sneden, C., Burles, S., Ivans, I. I., Beers, T. C., Truran, J. 
W., Lawler, J. E., Primas, F., Fuller, G. M., Pfeiffer, B., {\&} Kratz, 
K.-L. 2002, ApJ, 572, 861

Danzmann K., {\&} Kock M. 1982, J. Opt. Soc. Am., 72, 1556

Delbouille, L, Roland, G., {\&} Neven, L. 1973, Photometric Atlas of the 
Solar Spectrum from lambda 3000 to lambda 10000, (Li\`{e}ge, Inst. d'Ap., 
Univ. de Li\`{e}ge)

Den Hartog, E. A., Lawler, J. E., Sneden, C., {\&} Cowan, J. J. 2003, ApJS, 
148, 543

Den Hartog E. A., Lawler J. E., Sneden C., {\&} Cowan J. J. 2006, ApJS, 167, 
292

Edl\'{e}n, B. 1953, J. Opt. Soc. Am., 43, 339

Frebel A., Christlieb N., Norris J. E., Thom C., Beers T. C., {\&} Rhee J. 
2007, ApJ, 660, L117

Grevesse, N., {\&} Sauval, A. J. 1998, Space Sci. Rev., 85, 161

Grevesse, N., {\&} Sauval, A. J. 1999, A{\&}A, 347, 348

Grevesse, N., {\&} Sauval, A. J. 2002, Adv. Space. Res., 30, 3

Grevesse, N., Asplund, M., {\&} Sauval, A. J. 2007, Space Sci. Rev., 130, 
105

Grigoriev, I. S., {\&} Melikhov, E. Z. 1997, Handbook of Physical 
Quantities, (Boca Raton, CRC Press) p. 516

Hannaford, P., Lowe, R. M., Grevesse, N., Biemont, E., {\&} Whaling, W. 
1982, ApJ, 261, 736

Hashiguchi, S., {\&} Hasikuni, M. 1985, J. Phys. Soc. Japan 54, 1290

Hill, V., et al. 2002, A{\&}A, 387, 560

Holweger, H., {\&} M\"{u}ller, E. A. 1974, Sol. Phys., 39, 19

Irwin, A. W. 1981, ApJS, 45, 621

Ivans, I. I., Simmerer, J., Sneden, C., Lawler, J. E., Cowan, J. J., 
Gallino, R., {\&} Bisterzo, S. 2006, ApJ, 645, 613

Ivans, I. I., Sneden, C., Gallino, R., Cowan, J. J., {\&} Preston, G. W. 
2005, ApJ, 627, L145

Kurucz, R. L. 1998, in Fundamental Stellar Properties: The Interaction 
between Observation and Theory, IAU Symp. 189, ed T. R. Bedding, A. J. Booth 
and J. Davis (Dordrecht: Kluwer), p. 217

Kurucz R. L. 2007, 7. Linelists (\underline {http://kurucz.harvard.edu/})

Lawler, J. E., Bonvallet, G., {\&} Sneden, C. 2001a, ApJ, 556, 452

Lawler, J. E., Wickliffe, M. E., Den Hartog, E. A., {\&} Sneden, C. 2001b, 
ApJ, 563, 1075

Lawler, J. E., Wickliffe, M. E., Cowley, C. R., {\&} Sneden, C. 2001c, ApJS, 
137, 341

Lawler, J. E., Sneden, C., {\&} Cowan, J. J. 2004, ApJ, 604, 850

Lawler J. E., Den Hartog E. A., Sneden C., {\&} Cowan J. J. 2006, ApJS, 162, 
227

Lawler J. E., Den Hartog E. A., Labby Z. E., Sneden C., Cowan J. J., {\&} 
Ivans I. I. 2007, ApJS, 169, 120

Lodders, K. 2003, ApJ, 591, 1220

Malcheva, G., Blagoev, K., Mayo, R., Ortiz, M., Xu, H. L., Svanberg, S., 
Quinet, P., {\&} Bi\'{e}mont, E. 2006, MNRAS, 367, 754

Martin, W.C., Zalubas, R., {\&} Hagan, L. 1978, Atomic Energy Levels The 
Rare Earth Elements, NSRDS NBS 60 (Washington: U. S. G. P. O.) p. 174

Martin, W. C., Sugar, J., {\&} Musgrove, A. 2000, NIST Atomic Spectra 
Database, (http:// physics.nist.gov/PhysRefData/ASD/index.html)

Moore, C. E., Minnaert, M. G. J., {\&} Houtgast, J. 1966, The Solar Spectrum 
2934 {\AA} to 8770 {\AA}, NBS Monograph 61 (Washington: U.S. G. P. O.)

Musiol K., {\&} Labuz S. 1983, Physica Scripta 27, 422

Palmeri, P., Quinet, P., Wyart, J.-F., {\&} Bi\'{e}mont, E. 2000, Physica 
Scripta, 61, 323

Rosman, K. J. R., {\&} Taylor, P. D. P. 1998, Pure Appl. Chem., 70, 217

Smith V. V., Cunha K., {\&} Lambert D. L. 1995, AJ, 110, 2827 

Sneden, C. 1973, ApJ, 184, 839

Sneden C., Basri G., Boesgarrd A. M., Brown J. A., Carney B. W., Kraft R. 
P., Smith V., {\&} Suntzeff N. B. 1995, Publ. of the Astron. Soc. of the 
Pacific 107, 997 

Sneden C., McWilliam A., Preston G. W., Cowan J. J., Burris D. L., {\&} 
Armosky B. J. 1996, ApJ, 467, 819 

Sneden C., Cowan J. J., Ivans I. I., Fuller G., S. Burles, T. C. Beers, and 
J. E. Lawler 2000, ApJ, 533, L139 

Sneden C., Cowan J. J., {\&} Lawler J. E. 2003, Nuclear Phys. A718, 29c

Sneden, C., Cowan, J. J., {\&} Gallino, R. 2008, ARAA, in press

Stockett M. H., Den Hartog E. A., and Lawler J. E., 2007, J. Phys. B: 
Atomic, Mol., {\&} Opt. Phys. 40, 4529 

Xu H., Jiang Z., Zhang Z., Dai A., Svanberg S., Quinet P., {\&} Bi\'{e}mont 
E. 2003, J. Phys. B: Atomic, Mol., {\&} Opt. Phys. 36, 1771 

Xu, H. L., Jiang, H. M., Liu, Q., Jiang Z. K., {\&} Svanberg, S. 2004, Chin. 
Phys. Lett., 21, 1720

Westin, J., Sneden, C., Gustafsson, B., {\&} Cowan, J.J. 2000, ApJ, 530, 783

Whaling W., Carle M. T., {\&} Pitt M. L. 1993, J. Quant. Spectrosc. Radiat. 
Transfer 50, 7

Wickliffe, M. E., Lawler, J. E., {\&} Nave, G. 2000, J. Quant. Spectrosc. 
Radiat. Transfer, 66, 363

Woolf V. M., Tomkin J., {\&} Lambert D. L. 1995, ApJ, 453, 660 

\newpage 
\begin{center}
FIGURE CAPTIONS
\end{center}

Figure 1: Partial Grotrian diagram for Er \textsc{ii}. Upper and lower 
levels of both parities included in this study are shown.

Figure 2: FTS data from spectra {\#}7 of Table 1. The Er \textsc{ii} line 
near the center of the plot is from the odd-parity upper level at 28361.386 
cm$^{-1}$ to the even-parity ground level at 0.000 cm$^{-1}$. This UV line 
at 3524.913 {\AA} is the second strongest branch from the upper level with a 
branching fraction of 0.389 . There are Er \textsc{i }lines visible at a 
somewhat lower wavenumber and at a higher wavenumber near the left edge of 
the plot. Ringing from the apodization of the interferogram is visible as 
well as some weak isotopic structure near the base of the line.

Figure 3: FTS data from spectra {\#}6 of Table 1. The Er \textsc{ii} line 
near the center of the plot is from the odd-parity upper level at 28361.386 
cm-1 to the even-parity lower level at 21894.055 cm$^{-1}$. This IR line at 
15458.105 {\AA} is the second weakest branch reported from the upper level 
with a branching fraction of 0.00122 . The triplet structure is from the 
dominant even (nuclear spin I = 0) isotopes. 

Figure 4: Comparison of experimental transition probabilities from Musiol 
and Labuz (1983) to our transition probabilities as function of our 
transition probability or log(\textit{gf}), wavelength, and upper level energy.

Figure 5: Comparison of theoretical transition probabilities from Xu et al. 
(2003) to our transition probabilities as function of our transition 
probability or log(\textit{gf}), wavelength, and upper level energy.

Figure 6: Comparison of theoretical transition probabilities from Kurucz 
(2007) to our transition probabilities as function of our transition 
probability or log(\textit{gf}), wavelength, and upper level energy.

Figure 7: Relative transition strength factors, STR $\equiv $ 
log($\varepsilon $\textit{gf}) -- $\theta \chi $, for lines of Gd \textsc{ii} (Den 
Hartog et al. 2006) and Er \textsc{ii} (this study). For display purposes 
the long-wavelength limit has been set to 6000 {\AA}, which cuts out only 
some extremely weak lines of Gd \textsc{ii} and Er \textsc{ii} that can be 
detected neither in the Sun nor nearly all other stars. The short-wavelength 
limit of 2900 {\AA} covers all lines at that end of the spectrum in these 
two studies. Definitions of ``detection limit'' and ``strong lines'' of 
these species are given in the text.

Figure 8: Line-by-line Er abundances for the Sun and the $r-$process-rich 
metal-poor giant stars CS 22892-052, BD+17$^{o}$3248, HD 221170, HD 115444, 
and CS 31082-001, plotted as a function of wavelength. For each star, a 
dotted line is drawn at the mean abundance. As indicated in the figure 
legend, the three numbers in parentheses beside each star name are the mean 
abundance, the sample standard deviation $\sigma $, and the number of lines 
used in the analysis. The small scatter with an increased number of lines in 
comparison to earlier work (see text) yields improved accuracy and precision 
of abundance values.

Figure 9: Correlation of solar-system meteoritic and photospheric abundances 
for rare-earth elements studied in this series. The meteoritic abundances 
and their error estimates are those recommended by Lodders (2003). The 
sources of the photospheric abundances are: La, Lawler et al. (2001a); Nd, 
Den Hartog et al. (2003); Sm, Lawler et al. (2006); Eu, Lawler et al. 
(2001b); Gd, Den Hartog et al. (2006); Tb, Lawler et al. (2001c);

Ho, Lawler et al. (2004); Er, this study; and Hf, Lawler et al. (2007). 
Error bars adopted for the photospheric abundances are the sample standard 
deviations reported in those papers, which should be consulted for more 
detailed assessments. The dotted line indicates equality of the meteoritic 
and photospheric values. Note that a recently proposed renormalization by 
Grevesse, Asplund, {\&} Sauval (2007) would decrease the meteoritic 
abundances uniformly by 0.03 dex.

Figure 10: Comparison of rare-earth abundances in four $r-$process-rich stars to 
the solar-system $r-$process-only abundances. The solar-system values are taken 
from Simmerer et al. (2004). The stars, identified in the figure legend, are 
those that have been analyzed in this series of papers. For each star, the 
abundance differences have been normalized such that $\Delta $(log 
$\varepsilon $(Eu)) = 0. The dotted line indicates equality between the 
stellar and solar-system $r-$only abundances. The error bars are the means of 
the sigma values of individual stellar abundances. The abundances for the 
named elements are taken from the present and earlier papers of this series, 
and other abundances are taken from the original stellar analyses of the 
stars.

\end{document}